\documentclass[10pt,letterpaper,twocolumn,prl,showpacs]{revtex4-1}

\usepackage[english]{babel}

\usepackage[T1]{fontenc}
\usepackage{times,mathptmx}

\usepackage{amsmath,amssymb}
\usepackage[dvips]{graphicx,color}

\begin{document}

\title{Experimental observation of the spontaneous breaking of the time-reversal symmetry\\ in a synchronously-pumped passive Kerr resonator}

\author{Yiqing Xu}
\author{St\'ephane Coen}

\affiliation{Physics Department, The University of Auckland, Private Bag 92019, Auckland 1142, New Zealand}

\begin{abstract}%
  We experimentally observe a spontaneous temporal symmetry breaking instability in a coherently-driven passive
  optical Kerr resonator. The cavity is synchronously pumped by time-symmetric pulses yet we report output pulses
  with strongly asymmetric temporal and spectral intensity profiles, with up to 71\,\% of the energy on the same
  side of the pump center frequency. The instability occurs above a certain pump power threshold but remarkably
  vanishes above a second threshold, in excellent agreement with theory. We also observe a generalized bistability
  in which an asymmetric output state coexists with a symmetric one in the same pumping conditions.
\end{abstract}

\pacs{11.30.Qc, 05.45.-a, 42.60.Da, 42.65.Pc, 42.65.Sf, 42.65.Wi}

\maketitle

\raggedbottom

\noindent Spontaneous symmetry breaking (SSB) is a ubiquitous phenomenon in nature \cite{baker_spontaneous_1962,
*anderson_more_1972, *coleman_aspects_1988, *nambu_nobel_2009, malomed_spontaneous_2013}. It characterizes systems
described by equations invariant under certain symmetries but with lowest energy output states that do not exhibit
those symmetries. SSB is at the basis of many phase transitions and collective behaviors and accounts for a wide
range of fundamental effects including ferromagnetism, superconductivity, and convection cells
\cite{landau_statistical_1996, nicolis_self-organization_1977}. Its role is also famously recognized in cosmology
and particle physics, since it is tied to the celebrated Brout-Englert-Higgs mechanism \cite{englert_broken_1964,
*higgs_broken_1964}.

SSB occurs generically in many nonlinear Hamiltonian systems \cite{strocchi_symmetry_2008,
malomed_spontaneous_2013}. Cold interacting gases and nonlinear propagation of light are two examples of such
systems which have been thoroughly investigated in recent times. Both are described by various forms of the
nonlinear Schr\"odinger equation (NLSE) [also known as the Gross-Pitaevskii equation for Bose-Einstein condensates
(BECs)]. A plethora of configurations exhibiting SSB have been theoretically identified in these contexts, including
multicomponent BECs and solitons \cite{kockaert_stability_1999, esry_spontaneous_1999, coen_domain_2001}, nonlinear
couplers \cite{peschel_bistability_1994}, coupled microcavities \cite{maes_switching_2006}, or photonic lattices
\cite{de_sterke_spontaneous_2013}. SSB has also recently been proposed to realize time crystals
\cite{shapere_classical_2012}. Experimental demonstrations are less numerous but SSB has been observed unequivocally
both in BECs \cite{albiez_direct_2005} and in nonlinear optics \cite{yabuzaki_optical_1984,
cambournac_symmetry-breaking_2002, kevrekidis_spontaneous_2005, delque_symmetry-breaking_2011}.

SSB also occurs in dissipative systems. A famous example is the emergence of dissipative structures in chemical and
biological systems as described by Prigogine \cite{prigogine_symmetrybreaking_1967, *prigogine_symmetry_1969,
nicolis_self-organization_1977}. Remarkably dissipation may sometimes be itself responsible for breaking the
symmetry of the problem, which could have applications, e.g., for Brownian motors \cite{flach_directed_2000,
*gommers_dissipation-induced_2005}. In nonlinear optics, there is to our knowledge only one experimental
demonstration of SSB in a dissipative system, involving intracavity second-harmonic generation
\cite{longchambon_experimental_2005} although several other configurations have been studied theoretically
\cite{haelterman_pitchfork_1990, haelterman_symmetry-breaking_1995, *garcia-mateos_optical_1995}. The case of the
synchronously-pumped passive optical resonator filled with a Kerr nonlinear material studied in
Ref.~\onlinecite{haelterman_symmetry-breaking_1995, *garcia-mateos_optical_1995} is particularly interesting: it
exhibits a \textit{temporal} SSB instability in which the discrete time-reversal symmetry is broken. In contrast,
all reported experimental demonstrations of SSB in nonlinear optics are associated with asymmetries either in the
spatial (transverse) distribution of the light \cite{cambournac_symmetry-breaking_2002, kevrekidis_spontaneous_2005,
delque_symmetry-breaking_2011} or in its state of polarization \cite{yabuzaki_optical_1984,
longchambon_experimental_2005}. Remarkably, the SSB of passive Kerr resonators has been further reported twice
\cite{torres_bilateral_1999, schmidberger_spontaneous_2013} since its original prediction in 1995, yet has never
been studied experimentally. In this Letter, we report the first experimental observation of a time-reversal SSB
instability using an optical fiber ring cavity.

\begin{figure}[t]
    \centering
    \includegraphics[width=\columnwidth]{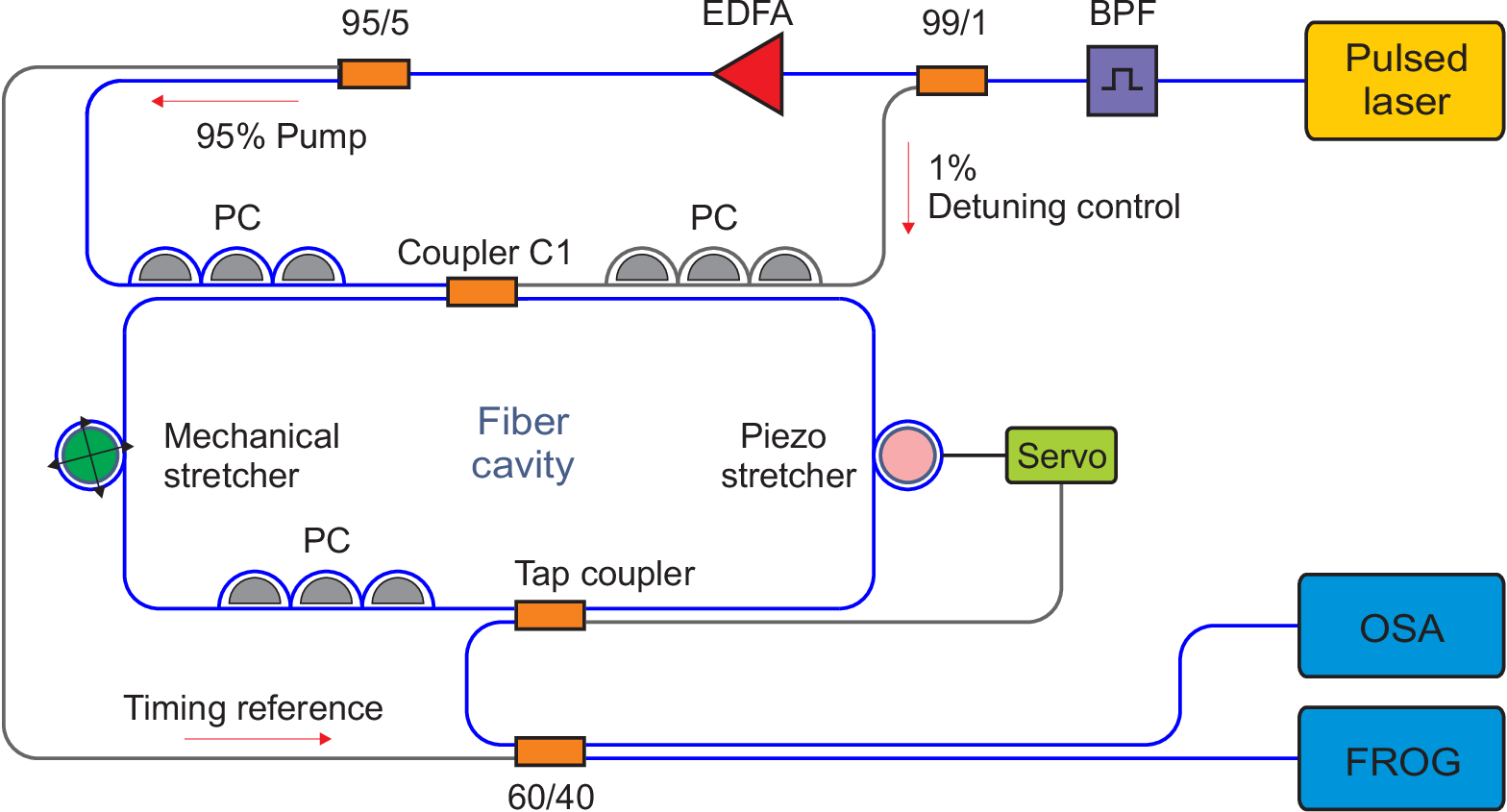}
    \caption{\small (Color online) Experimental setup: BPF: bandpass filter; EDFA: Erbium-doped fiber amplifier;
      PC, polarization controller.}
    \label{fig:setup}
\end{figure}

The cavity that we consider is depicted schematically in Fig.~\ref{fig:setup}. It is made up of single-mode silica
fiber that exhibits a nearly instantaneous self-focusing Kerr nonlinearity. The single-mode character of the
waveguide enables observations unhampered by the transverse diffraction of the beams and reduces the dynamics to a
single (temporal) dimension. Through fiber coupler C1, short pulses of light are launched into the cavity fiber,
where they propagate clockwise and experience chromatic dispersion and nonlinearity. At each round trip in the
cavity, some light is lost but, more importantly, at each pass in coupler C1 the external pump pulses are added
coherently and synchronously to the intracavity light wave. All these effects can be described by a single partial
differential equation resulting from an averaging procedure that involves the NLSE and the cavity boundary
conditions \cite{haelterman_additive-modulation-instability_1992},
\begin{equation}
    \frac{\partial E(z,\tau)}{\partial z} = \left[-1 + i(|E|^2 - \Delta) - i\eta\frac{\partial^2}{\partial\tau^2}\right] E +
    S(\tau).\label{eq:LL}
\end{equation}
This mean-field equation is identical to the Lugiato-Lefever model of a diffractive cavity
\cite{lugiato_spatial_1987}. It is valid provided the field evolves slowly enough over one cavity round-trip, i.e.,
for cavities of high enough finesse. Here the variable $z$ describes the slow evolution of the intracavity field~$E$
over successive round-trips and is normalized to the photon mean-free path in the cavity while $\tau$ is a fast-time
defined in a reference frame that travels at the group velocity of light in the fiber and allows for the description
of the temporal profile of the intracavity pulse envelope. The terms on the right-hand side of Eq.~(\ref{eq:LL})
account for, respectively, cavity losses, Kerr nonlinearity, cavity phase detuning, chromatic dispersion, as well as
external pumping. With our normalization \cite{leo_temporal_2010, *jang_ultraweak_2013}, the cavity phase detuning
$\Delta=\delta_0/\alpha$, where $\alpha$ is half the fraction of power lost per round-trip (coupler transmission
plus fiber absorption; finesse $\mathcal{F}=\pi/\alpha$), and $\delta_0 = 2m\pi - \phi_0$, with $\phi_0$ the overall
cavity round-trip phase shift and $m$ the order of the closest cavity resonance. $\eta$ is the sign of the
group-velocity dispersion coefficient $\beta_2$ of the fiber. We take $\eta=-1$ as with a self-focusing nonlinearity
the temporal SSB instability of the passive Kerr cavity only occurs with anomalous dispersion
\cite{haelterman_symmetry-breaking_1995, *garcia-mateos_optical_1995}. Finally, $S(\tau)$ is the field envelope of
the external pump pulses.

\begin{figure}[t]
    \centering
    \includegraphics[width=\columnwidth]{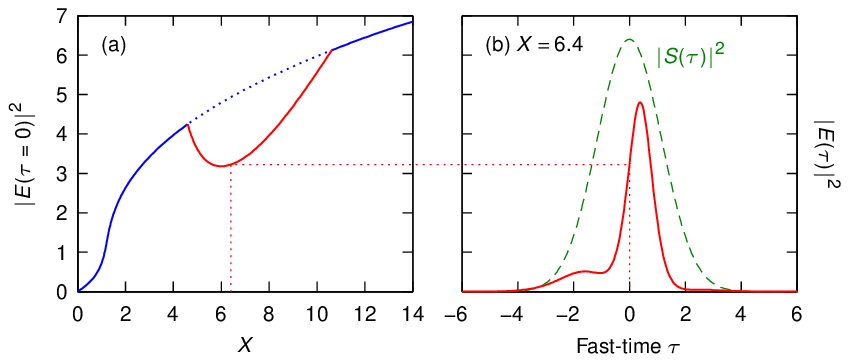}
    \caption{\small (Color online) (a) SSB bifurcation diagram with blue (red) curves representing the
      symmetric (asymmetric) states, respectively (dotted parts are unstable), for $\Delta=0.92$ and $T_0=2.3$. (b)
      Asymmetric pulse temporal intensity profile obtained for $X=6.4$ (red) superimposed with the pump
      pulse (dashed green).}
    \label{fig:bifurc}
\end{figure}

Let us first briefly describe the SSB instability of the passive Kerr cavity with the help of numerical simulations.
We are interested in the temporal envelope of the intracavity pulses under steady-state conditions ($\partial
E/\partial z=0$). These were computed by looking for the zeros of the right-hand side of Eq.~(\ref{eq:LL}) with a
generalized Newton-Raphson solver and a continuation method. Provided the pump pulse field profile is temporally
symmetric, $S(\tau) = S(-\tau)$, the mean-field model Eq.~(\ref{eq:LL}) is symmetric under a time reversal
transformation, $\tau \rightarrow -\tau$, yet it admits asymmetric solutions. This behavior is illustrated in
Fig.~\ref{fig:bifurc} obtained for a detuning $\Delta=0.92$ and for symmetric chirp-free Gaussian pump pulses
$S(\tau) = \sqrt{X}\exp[-(\tau/T_0)^2]$, with $T_0=2.3$. Figure~\ref{fig:bifurc}(a) represents the bifurcation
diagram of the SSB instability where we have plotted the power of the intracavity pulses at $\tau=0$, i.e., at a
time corresponding to the center of the pump pulses, versus the pump peak power~$X$. For low pump power, the
intracavity pulses are temporally symmetric (blue curve). However, above a certain pump power threshold, the
symmetric state becomes unstable (dotted part) and is replaced by an asymmetric one (red curve). The corresponding
drop in $|E(\tau=0)|^2$ observed at that point is associated with the peak of the intracavity pulse shifting to the
side and breaking its alignment with the center of the pump pulse. This is illustrated in Fig.~\ref{fig:bifurc}(b)
where we have plotted the temporal intensity profile of the pump pulse (dashed green) superimposed with the
asymmetric intracavity pulse obtained for $X=6.4$ (red). Of course its mirror image (not shown for clarity) is also
a solution of the problem. We must point out that the temporal duration of the main peak of the asymmetric
intracavity pulse is set by a balance between nonlinearity and dispersion, and is comparable with the duration of
the cavity solitons that the Kerr cavity is known to support \cite{leo_temporal_2010, *jang_ultraweak_2013}. With
wider pump pulses (larger values of $T_0$) than those considered here, we would observe intracavity pulses that
break up into multiple peaks (modulational instability) of similar duration
\cite{haelterman_additive-modulation-instability_1992, coen_competition_1999, *coen_continuous-wave_2001}. While SSB
manifests itself even in that regime, the experimental signature of the asymmetry is less pronounced, hence we have
chosen to work here with values of $T_0$ of the order of unity. Finally, let us also note in
Fig.~\ref{fig:bifurc}(a) that the symmetry of the solution is restored above a second pump power threshold.

To experimentally observe the SSB instability described above, we have used pulses of $2.3$~ps duration (full-width
at half maximum) at 20~MHz repetition rate as the pump. These pulses were obtained by temporally-broadening the
output of a mode-locked fiber laser with a bandpass filter (see Fig.~\ref{fig:setup}) and were checked to have a
nearly chirp-free Gaussian profile using frequency resolved optical gating (FROG). The ring cavity was made up of
standard telecommunication fiber which at the 1550~nm pump laser wavelength exhibits chromatic dispersion and
nonlinearity coefficients $\beta_2 = -21\ \mathrm{ps^2/km}$ and $\gamma = 1.2\ \mathrm{W^{-1}\,km^{-1}}$,
respectively. The cavity length $L$ was set to about $10.3$~m to realize the condition of synchronous pumping, i.e.,
to match the time-of-flight of the light pulses in the cavity with the pump laser repetition time ($1/20\
\mathrm{MHz}=50$~ns). Fine adjustment was performed with a mechanical fiber stretcher driven by a translation stage
with an accuracy of about 20~fs. The pump pulses were amplified up to a maximum peak power $P_\mathrm{in,peak}$ of
about~20~W before being launched into the cavity through coupler C1, which had a transmission coefficient
$\theta_1=20$\,\%. A 1\,\% tap coupler placed inside the cavity was used to monitor the intracavity field with an
optical spectrum analyzer (OSA) and a FROG apparatus. Note that a fraction of the pump beam was directly combined
with that output to serve as a timing reference and to lift the FROG time-direction ambiguity. The cavity finesse
was measured to be about~20 ($\alpha=0.15$). With these notations, the normalized pump peak power $X=\gamma L
P_\mathrm{in,peak} \theta_1/\alpha^3$ (it can range from 0 to~15), and the normalized pump pulse width $T_0$ is
found to be about $2.3$, as used in our numerical simulations. Finally, let us point out that the cavity
detuning~$\Delta$ was controlled, measured, and stabilized with a servo system driven by a small fraction of the
pump laser launched in the cavity in the opposite direction as described in detail in
Ref.~\cite{coen_experimental_1998}.

\begin{figure}[t]
    \centering
    \includegraphics[width=\columnwidth]{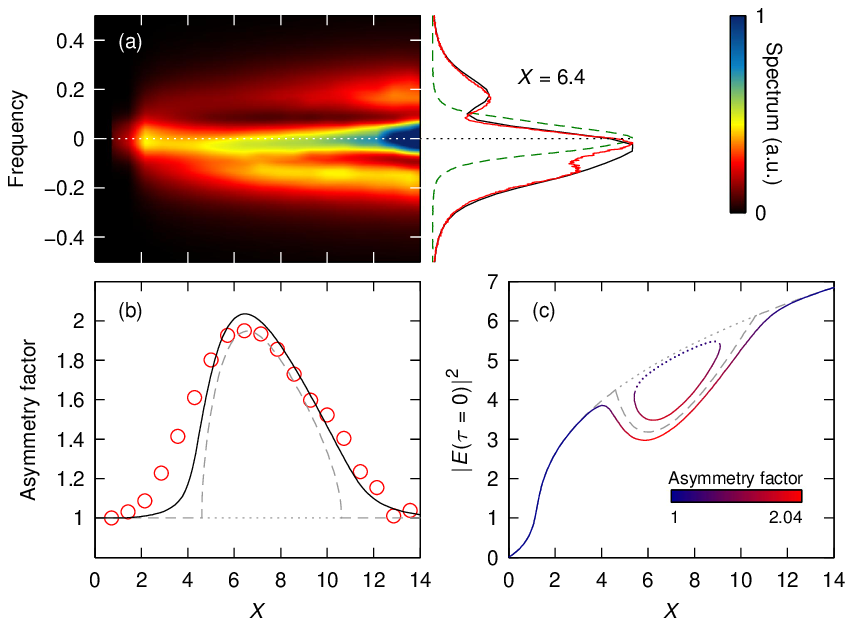}
    \caption{\small (Color online) (a) Experimental output spectra (linear scale) for $\Delta=0.92$ versus normalized
    pump power~$X$. Right: spectrum obtained for $X=6.4$ (red) compared with the simulated one (black) and
    superimposed on the measured pump spectrum (dashed green). (b) Spectral asymmetry factor derived from (a) [red
    circles] and compared to numerical results without (light gray) and with (black) the $\beta_3$ contribution.
    (c) Bifurcation diagram as in Fig.~\ref{fig:bifurc}(a) [light gray curves] compared to that inclusive of the
    $\beta_3$ contribution (color-coded with the asymmetry factor). Dotted parts correspond to unstable states.}
    \label{fig:exp1}
\end{figure}

As temporally asymmetric pulses are usually associated with asymmetric spectral densities, we relied on spectral
measurements of the cavity output as a convenient signature of the breaking of the time-reversal symmetry in our
experiment. A first set of experimental measurements is presented in Fig.~\ref{fig:exp1} and was obtained with a
detuning $\Delta=0.92$ identical to that used for the numerical results of Fig.~\ref{fig:bifurc}. Output spectra
were measured for monotonically increasing values of the pump peak power~$X$ and are shown as a pseudocolor plot in
Fig.~\ref{fig:exp1}(a). A clear shift of the spectral power towards the low-frequency components can be readily
observed as the pump power increases. To highlight this aspect, the spectrum obtained for $X=6.4$ is plotted along
the right side of Fig.~\ref{fig:exp1}(a) (red) superimposed with the measured pump spectrum (dashed green) for
comparison. Nearly 66\,\% of the intracavity spectral power is located on the low frequency side of the pump center
frequency. A good agreement is observed with the theoretically modeled spectrum (black), which corresponds to that
of the pulse illustrated in Fig.~\ref{fig:bifurc}(b).

For a more comprehensive analysis of the observed symmetry breaking bifurcation, we define the spectral asymmetry
factor as the ratio of the integrated spectral density on the low frequency side of the pump center frequency to
that on the high frequency side. The asymmetry factor can be simply evaluated from the data shown in
Fig.~\ref{fig:exp1}(a) and the result is plotted in Fig.~\ref{fig:exp1}(b) [red circles]. At low pump power, it is
initially equal to~1 (corresponding to a symmetric state) but rapidly rises to a maximum of about $1.9$ obtained for
a normalized pump power $X = 6.4$ and for which we observe the maximum asymmetry for this detuning. Comparison with
numerical simulations (dashed light-gray) reveals a reasonable quantitative agreement. We note however that while
the theory predicts a rather abrupt onset and disappearance of symmetry breaking, the experimentally observed
transitions are much more progressive. We have traced this difference to the presence of a small amount of
third-order dispersion in the cavity fiber with coefficient $\beta_3 = 0.1\ \mathrm{ps^3/km}$. This weakly breaks
the perfect time-reversal symmetry of the problem \cite{leo_nonlinear_2013}, and introduces an additional term $d_3
\partial^3 E/\partial\tau^3$ on the right-hand side of Eq.~(\ref{eq:LL}), with $d_3 =
\sqrt{2\alpha/L}\,\beta_3/(3|\beta_2|^{3/2}) \simeq 0.002$. As can be seen in Fig.~\ref{fig:exp1}(b), the asymmetry
factors calculated with this small contribution taken into account (black) better fit the experimental data. In
Fig.~\ref{fig:exp1}(c), we also show how the SSB bifurcation diagram changes in presence of third-order dispersion.
Here the light gray curves are identical to those shown in Fig.~\ref{fig:bifurc}(a). The curves inclusive of the
small $\beta_3$ contribution are color-coded as a function of the asymmetry factor to highlight that a clear
distinction still exists between the ``symmetric'' state (blue) and the asymmetric ones. The major difference is the
lifting of the degeneracy between the two original mirror-like asymmetric solutions (which now sit on two
disconnected curves) and this is associated with the change from an abrupt to a progressive symmetry breaking
transition. In these conditions, ramping up the pump power should allow only one enantiomer to be observed. However,
the system is very sensitive to a small cavity synchronisation mismatch or equivalently to fluctuations in the pump
laser repetition rate \cite{coen_convection_1999}, and in practice we could readily observe both left- and
right-handed asymmetric states in the experiment. We must also stress out that the decay of the asymmetry factor
past the maximum [see Fig.~\ref{fig:exp1}(b)] and the restoration of nearly perfect symmetric conditions above $X
\sim 11$ is remarkable in this context. It is as expected from the theory [Fig.~\ref{fig:bifurc}(a)] and confirms
that the observed asymmetry is tied to the temporal SSB instability of the nonlinear cavity dynamics.

\begin{figure}
    \centering
    \includegraphics[width=\columnwidth]{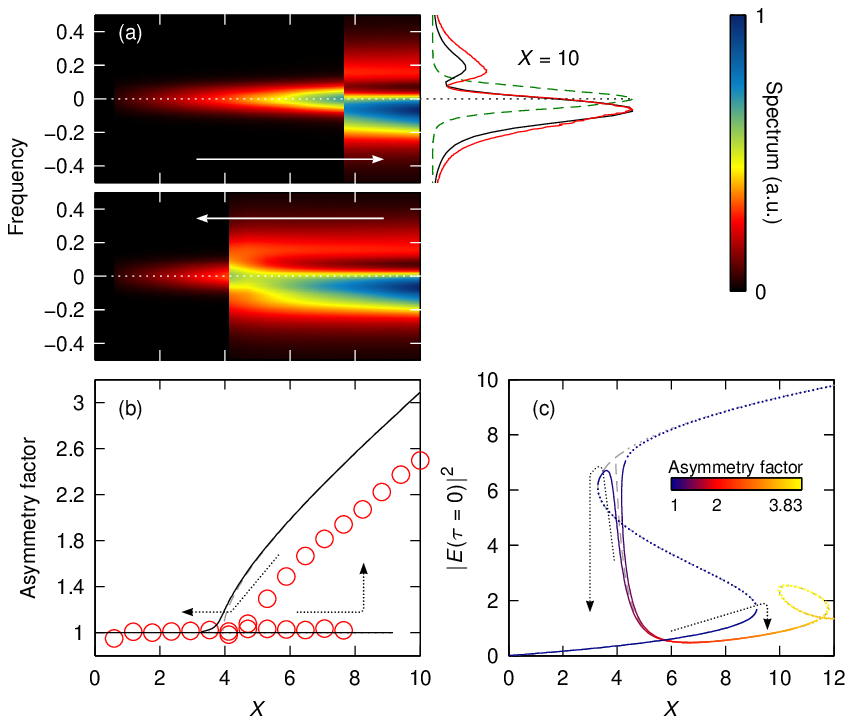}
    \caption{\small (Color online) As in Fig.~\ref{fig:exp1} but for $\Delta = 3.2$. In (a) the top (bottom) panel
    was obtained for increasing (decreasing) values of $X$, respectively. The spectrum on the right corresponds
    to $X=10$. In (b)--(c), arrows indicate how the hysteresis is traveled through.}
    \label{fig:exp2}
\end{figure}

A second set of measurements is presented in Fig.~\ref{fig:exp2} for a higher detuning, $\Delta=3.2$. In this regime
the Kerr cavity is known to exhibit hysteresis ($\Delta>\sqrt{3}$, see Refs.~\cite{haelterman_dissipative_1992,
coen_experimental_1998}). This is clearly evidenced in our recorded spectra [Fig.~\ref{fig:exp2}(a)], where the top
(bottom) panel was obtained for monotonically increasing (respectively, decreasing) values of the pump peak
power~$X$. A clear hysteresis is also seen in the spectral asymmetry factors deduced from the measured spectra
[Fig.~\ref{fig:exp2}(b)]. Theory (black) and measurements (red circles) are again in good qualitative agreement. For
increasing values of $X$, we start on the lower symmetric branch (asymmetry factor of~1). At the end of this branch
($X \simeq 8$), the corresponding upper symmetric state is unstable as highlighted by the bifurcation diagram
[Fig.~\ref{fig:exp2}(c)] so the cavity abruptly switches to a stable \emph{asymmetric} state instead. Here the
measured asymmetry factor jumps from 1 to~$\sim 2$. We observe a maximum asymmetry factor of $2.5$ for $X=10$ [i.e.,
$71\,\%$ of the spectral power is on the same side of the pump center frequency, see spectra highlighted on the
right of Fig.~\ref{fig:exp2}(a)], above which no more stable solutions can be observed and the intracavity pulse
breathes periodically. Upon decreasing $X$, the cavity remains in the asymmetric state until it smoothly reconnects
with the stable part of the \emph{upper} symmetric branch at $X \simeq 4.7$. When $X$ then goes below about $4.1$,
we see an abrupt change in the output spectrum (but no change in the asymmetry factor which stays equal to~1) as the
cavity switches from the upper symmetric state back to the lower symmetric state. From $X=4.1$ to $X=8$, we note the
coexistence of the stable symmetric lower branch solution with the asymmetric solution, i.e., generalized
bistability \cite{torres_bilateral_1999}.

\begin{figure}
    \centering
    \includegraphics[width=\columnwidth]{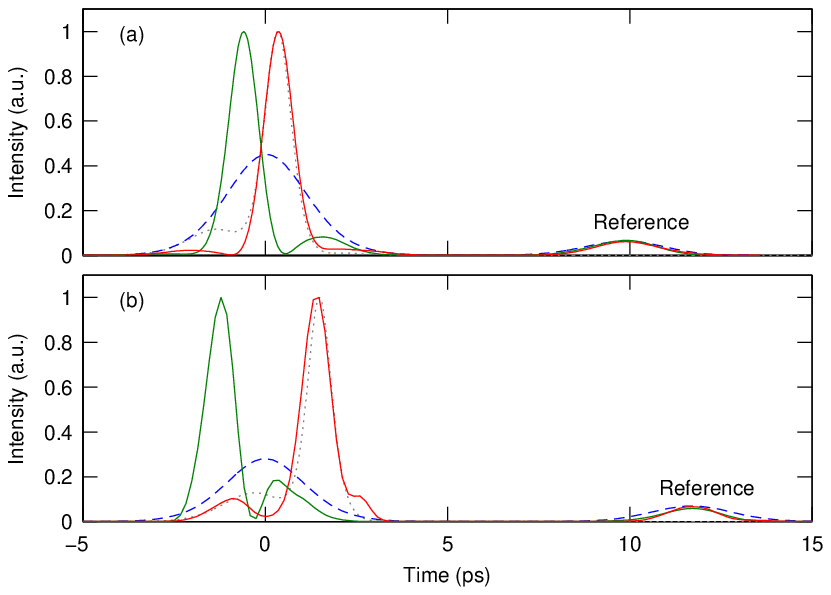}
    \caption{\small (Color online)
    Experimental temporal intensity profiles of the two mirror-like stationary intracavity pulses (red and green)
    retrieved with FROG for (a) $X = 6.4$, $\Delta = 0.92$ and (b) $X = 10$, $\Delta = 3.2$. The dashed curves are the
    symmetric pulses observed at low pump power level while the gray dotted curves are theoretical predictions.}
    \label{fig:FROG}
\end{figure}

To complete our study, we performed FROG measurements of the asymmetric states with the largest observed spectral
asymmetry, i.e., for $\Delta=0.92$, $X=6.4$ and $\Delta=3.2$, $X=10$. The retrieved temporal intensity profiles are
shown in Figs.~\ref{fig:FROG}(a) and (b), respectively. In each case, the two mirror-like solutions are shown. As
explained above we switch between the two through a small cavity desynchronisation of $\sim 20$~fs. In each figure,
the measured asymmetric pulses are superimposed with the temporal profile of a pulse in the symmetric state measured
at lower pump power (dashed curve). The two asymmetric pulses are not perfect mirror-image of each other, which we
attribute to the weak third-order dispersion and the non-perfect synchronization. Still, they have a strong
asymmetric temporal profile compared with the symmetric state. In particular, their peak is significantly temporally
shifted from the center, and the observed shift agrees remarkably well with theory as shown by the profiles obtained
from numerical simulations (dotted gray curves). Note that the delayed subpulses on the right of
Figs.~\ref{fig:FROG}(a)--(b) are the reference pulses which were split from the pump before entering the cavity and
recombined at the output, providing a timing reference for the FROG measurement. These were all aligned onto each
other when plotting the results.

To conclude, we have presented what we believe is the first experimental observation of the spontaneous breaking of
the time-reversal symmetry in a nonlinear dissipative system. Our study performed in a synchronously-pumped Kerr
optical fiber cavity has revealed the generation of pulses with a strongly asymmetric temporal intensity profile
from a seemingly time-symmetric configuration in excellent agreement with theory. A generalized bistability between
symmetric and asymmetric solutions has also been unequivocally observed. The passive Kerr resonator, with its
simplicity and ubiquitous character, is already considered as the paradigm of nonlinear systems subject to
instabilities. Given the important role of symmetries in physics, our study further reinforces this status. Our
observations may also have implications for other systems based on nonlinear cavities including microresonators and
in which temporal SSB may lead to uncontrollable timing jitter in the pulse trains generated by such devices.

\end{document}